\documentclass[%
 aip,
 jmp,%
 amsmath,amssymb,
 reprint,%
]{revtex4-1}

\usepackage{graphicx}
\usepackage{dcolumn}
\usepackage{bm}
\usepackage{lipsum}  
\usepackage{epstopdf}
\usepackage{xcolor}
\usepackage{tabularx}

\begin{document}


\title[]{Freestanding and flexible composites of magnetocaloric Gd$_5$(Si,Ge)$_4$ microparticles embedded in thermoplastic poly(methyl methacrylate) matrix}
\author{Vivian M. Andrade}
\affiliation{%
	'Gleb Wataghin' Physics Institute, State University of Campinas (IFGW-UNICAMP), Rua Sergio Buarque de Holanda, 777, 13083-859 Campinas - SP, Brazil 
}%
\affiliation{%
	IFIMUP and IN-Institute of Nanoscience and Nanotechnology, Departamento de F\'{i}sica da Faculdade de Ci\^{e}ncias, Universidade do Porto, Rua do Campo Alegre, 687, 4169-007 Porto, Portugal
}%

\author{Nathalie B. Barroca}

\author{Ana L. Pires}
\author{Jo\~{a}o H. Belo}
\author{Jo\~{a}o P. Ara\'{u}jo}
\author{Andr\'{e} M. Pereira}
\affiliation{%
	IFIMUP and IN-Institute of Nanoscience and Nanotechnology, Departamento de F\'{i}sica da Faculdade de Ci\^{e}ncias, Universidade do Porto, Rua do Campo Alegre, 687, 4169-007 Porto, Portugal
}%

\date{\today}

\begin{abstract}


The implementation of processed magnetic materials onto thermoplastics can be an approach for practical application of brittle intermetallic materials with the advantage of enlarging the range of applications. In the present work, we evaluate the effect of blending magnetocaloric Gd$_5$Si$_{2.4}$Ge$_{1.6}$ micrometric particles with 3.4 $\mu$m in different weight fractions onto a flexible, transparent and non-magnetic poly(methyl methacrylate) (PMMA). A close to homogeneous grain distribution along the polymer surface were achiever by using a simple solvent casting method for evaluation of their magnetocaloric properties. From XRD analysis, it was found a relative unit cell volume reduction of $\sim$2.5$\times$10$^3$ ppm for the composite with 70 wt.\% of powder as a result of interfacial interactions between the components. Although PMMA does not influence the magnetic nature of microparticles main phase, a reduction on the amount of secondary monoclinic phase occurs for all produced composite samples. As a consequence, a weakening on the effect of secondary phases on the micropowder magnetocaloric response is observed as a result of hydrostatic pressure from the difference between thermal expansions of matrix and filler.
\end{abstract}

\maketitle


\section{Introduction}

Magnetocaloric materials for heating/cooling applications at room temperature have been a field of interest for several decades for substitution of hazardous gases used in traditional vapour compression refrigerators. The discovery of a Giant Magnetocaloric Effect (GMCE) on the Gd$_5$Si$_2$Ge$_2$ compound\cite{pecharsky1997phase} was a milestone due to the potential for magnetic refrigeration at room temperature. Since then, several strategies have been followed in order to evaluate/increase the MCE properties of the Gd$_5$(Si,Ge)$_4$ family compounds aiming new applications, such as micro-cooling devices and medical treatments\cite{boekelheide2017gd,kitanovski2015overview,el2017ferromagnetic}. An intensive research on this family led to the observation of impressive effects that raise from their strong magnetostructural coupling, such as: large magnetoresistance and colossal magnetostriction\cite{pecharsky2001gd5,pires2014phase}. The crystallographic arrangement on these compounds is composed by a stack of pseudo-cubes connected by Si/Ge dimmers positioned at the cube edges, the so-called interslab positions, that are extremely sensitive to internal and external parameters. The number of Si/Ge bonds will determine the crystallographic structure and, consequently, the magnetic ordering: orthorhombic-II [O(II)] Gd$_5$Ge$_4$-type does not present any bond with an associated antiferromagnetic state, monoclinic (M) Gd$_5$Si$_2$Ge$_2$-type where half of this bond is formed and orthorhombic-I [O(I)] with all Si/Ge bonds with both presenting a ferromagnetic ordering\cite{pecharsky1997phase}. For this reason, thermal variations, applied magnetic fields and hydrostatic pressure, for example, can lead to the rupture or formation of Si/Ge dimmers and, thus, changing the structure\cite{gschneidnerjr2005recent}. However, most of the published research is performed with bulk samples that are brittle and fragile\cite{harstad2017enhancement,ozaydin2014multi}, which represents a significant drawback in shape design for technological applications. Besides that, when implemented into devices, the material is submitted to several thermal cycles that can lead to its degradation\cite{booth2012magnetocaloric,pires2015influence}. An approach to improve the material mechanical and chemical stability can be achieved by incorporating the inorganic magnetocaloric materials (MCM) into organic polymers with suitable properties. Particularly, flexibility on magnetic composites allows implementation on sensors and transducers\cite{akdogan2005piezoelectric}, dielectric wave-guide\cite{xiang2007dielectric}, microwave devices\cite{bora2019gadolinium} and bioengineering\cite{laurencin1999tissue}. 

Generally, there has been an increase on research activities in polymer-based composites on the last 20 years focused on implementation in several industries such as transport, military, aerospace\cite{koniuszewska2016application}, biomedical\cite{ramakrishna2001biomedical} and textile\cite{gereke2013experimental}. There are few reports dedicated to the studies of Gd-based materials in polymeric matrices. For instance, Ozaydin et al. have shown that the incorporation of low amounts ($<$ 10\%) of Gd$_5$Si$_2$Ge$_2$ micropowder in polyvinylidene fluoride (PVDF) by spin coating revealed an enhancement on converting magnetic energy into electrical power\cite{ozaydin2014multi}. More recently, Bora and co-workers demonstrated the potential of applying Gd$_5$Si$_4$ milled powder blended with the elastomer polydimethylsiloxane (PDMS) on microwave absorption in the Ku-band\cite{bora2019gadolinium}. Regarding the MCE studies, Imamura \textit{et. al.} have evaluated the effect of compacting and sintering Gd$_{5.09}$Si$_{2.03}$Ge$_{1.88}$ powder with low amounts ($\sim$ 15\%) on a camphorsulfonic acid doped polyaniline (PANI-CSA) conductive polymer\cite{imamura2017new}. The authors observed a $\sim$ 22\% reduction on the refrigerant cooling power (RCP) values and the gain on mechanical properties had no influence the magnetic nature of the powder. 


In this work, we have choose a more flexible poly(methyl methacrylate) thermoplastic to reinforce powders of Gd$_5$Si$_{2.4}$Ge$_{1.6}$ powder with 3.4 $\mu$m. The evaluation on the structural, magnetic and magnetocaloric effect of blending different weight fractions of the micropowder into a flexible, transparent and resistant PMMA using a simple and reproducible chemical technique. The chosen stoichiometry displays a broader working range temperature for the MCE at room temperature and with absence of hysteresis losses\cite{hadimani2015investigation,misra2006distribution}. The PMMA is suitable for making flexible magnetocaloric composites due to its large durability, flexibility, high resistance to scratches and very low water absorbency ($\sim$ 3\%)\cite{ali2015review,kim2010dissolvable,park2003relationship}. Considered as the hardest thermoplastic, PMMA presents a thermal stability in a wide range of temperature going from 200 K to 500 K which is of great importance for room temperature applications.  

\section{Experimental techniques}

\subsection{Bulk and powder synthesis}

The magnetocaloric filler for the composite solution were prepared through the Tri-arc technique under Ar atmosphere\cite{belo2012phase}. For this, each constituent elements (with purities higher than 99.99\%) were first accurately weighted in order to obtain the Gd$_5$Si$_{2.4}$Ge$_{1.6}$ composition. Since the weight loss, after arc-melting, was less than 2\%, the final product stoichiometry was assumed unchanged. No heat treatment was performed on the as-cast ingot that was manually grounded and sieved through several filters with hole sizes from 50 $\mu$m to 5 $\mu$m in order to obtain a thinner powder and to guarantee a homogeneous dispersion on the composite.

\subsection{Composites synthesis}

The polymer composite samples were prepared via solvent casting technique using Gd$_5$Si$_{2.4}$Ge$_{1.6}$ (GSG) 3.4 $\mu$ m particles obtained as described above\cite{marycz2016polyurethane}. The composite solutions were prepared by dissolving PMMA in dichloromethane (DCM) (Sigma, 270997-1L) at 40 $^0$C until complete dissolution was achieve in order to acquire a 10wt.\% PMMA blend. Afterwards, composite solutions with GSG weight fractions of 10, 30, 50 and 70\% were obtained by simply dispersing the GSG microparticles in the PMMA solution. The resultant solutions were then solvent casted in order to obtain the freestanding and flexible magnetocaloric films. Different moulds shapes were used to illustrate the possibility of producing flexible magnetocaloric composite materials with variable designs. Two of the products obtained through this technique are depicted in Fig. \ref{XRD}(a) and (b) where is possible to notice that the composites surface becomes visually tougher and darker as the GSG weight fraction in PMMA increases.

\subsection{Characterization techniques}

Crystallographic characterization was performed via X-ray diffraction (XRD) data obtained at room temperature using a Rigaku SmartLab diffractometer with Cu-K$\alpha$ radiation (1.540593 \AA{}), 45 kV and 200 mA at IFIMUP. The diffraction patterns were collected from $20^o \leq 2\theta \leq 60^o$ range in a Bragg Brentano geometry, with 0.02 $^o$ steps. Analysis were performed through Rietveld refinement considering a Pseudo-Voigt function of the XRD patterns using the FullProf software \cite{rodriguez1993recent}. Morphological properties of the composites were evaluated by Scanning Electron Microscopy (SEM) using a \textit{Philips-FEI/Quanta 400}. Cross-section imaging were performed on freeze fractured samples using Nitrogen. Superconducting Quantum Interference Device (SQUID) was used for the magnetic characterization and evaluation of the MCE on [5,350] K temperature range. 

\section{Experimental results}

\subsection{Crystallographic and morphological characterization}

\begin{figure}
	\centering
	\includegraphics[width=8.5cm]{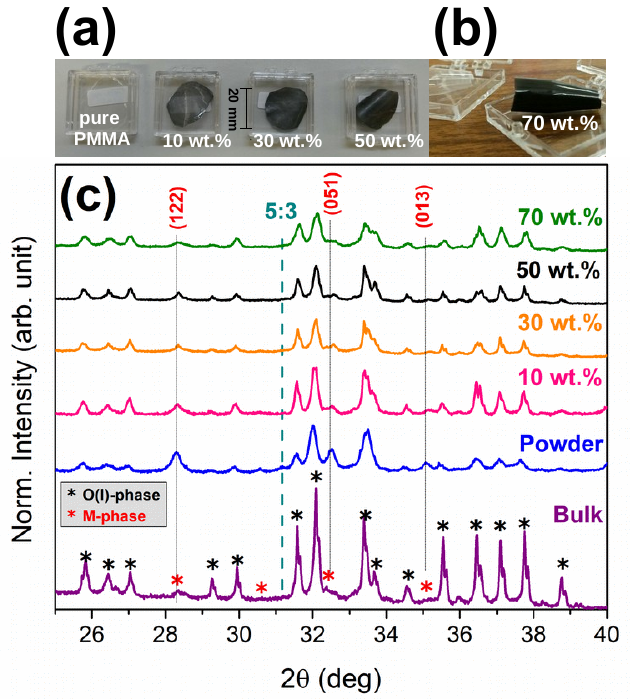}
	\caption{\textbf{(a)} Photographs of pure PMMA and the embedded with 10 wt.\%, 30 wt.\% and 50 wt.\% composites by using a flat mould and \textbf{(b)} of 70 wt.\% by using a simple eppendorf as a mould. \textbf{(c)} X-ray diffraction patterns for the composites with different concentrations, powder and bulk of Gd$_5$Si$_{2.4}$Ge$_{1.6}$ with the corresponding peak positions of the structures present on the samples: orthorhombic-I [O(I)], monoclinic (M) and R$_5$M$_3$ (5:3) phases returned from Rietveld calculations.}\label{XRD}
\end{figure}

The XRD patterns for bulk, powder and composites samples are presented in Fig. \ref{XRD}(c). At a first glance, one highlight is the broadening of the diffracted peaks on the used powder in comparison the bulk counterpart. This is a direct consequence on the reduction of particle size after sieving\cite{andrade2016magnetic,pires2015influence,hunagund2018investigating}. From \textit{Rietveld} refinements it was found the formation of expected O(I) structure, M-phase and eutectic R$_5$M$_3$-phase (that will be simply denoted as 5:3)\cite{belo2012phase}, with the results summarized for all samples on Table \ref{PMMA_parameters_XRD}. Misra \textit{et al.} have shown that at the Si-rich region of Gd$_5$(Si,Ge)$_4$ family compounds, Ge preferentially occupies the interslab positions at the O(I) structure and these values were considered for the calculations\cite{misra2006distribution}. As for the M-phase, the Si/Ge sites were initially considered to be fully occupied and, after the refinement, it was found the same preferential occupation of Ge atoms positioned at the shortest distance between the dimers, with the atomic positions also displayed in SID Table S3. As highlighted by vertical lines in Fig. \ref{XRD}(c), the corresponding peaks intensities for M phase increase for the micropowder, indicating an increment on this phase amount with decreasing particle size. Such observation is confirmed by \textit{Rietveld} refinement results which give the presence of more than 20\% of M-phase. Similarly, the amount of detected eutectic phase was found to increase for Gd$_5$Si$_4$ compound when the particle size reduces from 700 to 80 nm through ball milling\cite{hunagund2018investigating}. Previously, it was shown an enlargement of secondary R$_5$M$_4$ phases, such behavior was observed for milled Gd$_5$Si$_{1.3}$Ge$_{2.7}$ and Tb$_5$Si$_2$Ge$_2$ followed by a reduction on the unit cell volume of the main phase\cite{pires2015influence}. In our case, by sieving the GSG powders, a slight increase on the unit cell volume of the main O(I) phase from 865 \r{A}$^3$ for the bulk to 869 \r{A}$^3$ for the 3.4 $\mu$m powder was noted, while M and 5:3 unit cells remain unchanged, as summarized on Table \ref{PMMA_parameters_XRD}. With the selection of smaller particles by the strainers, the defects at the grain boundaries become more evident and can be leading to the detection of larger amounts of the distorted monoclinic phase\cite{toraya1995quantitative}. In another words, there is a self-segregation of the crystallographic structures on the Gd$_5$Si$_{2.4}$Ge$_{1.6}$ compound since there is no thermal effects involved on the chosen technique for grain separation\cite{pecharsky1997phase,belo2012phase}. 

\begin{table*}[]
	\centering
	\caption[Lattice parameters and fit values for \textit{Rietveld} refinement of GSG/PMMA composites.]{Returned lattice parameters, Pseudo-Voigt profile curve U, V, W values and goodness of fit for Gd$_5$Si$_{2.4}$Ge$_{1.6}$/PMMA composites \textit{Rietveld} refinements. (Note: the U,V,W values correspond to the main O(I)-phase).}\label{PMMA_parameters_XRD}
	\begin{tabular}{|c|c|c|c|c|c|c|c|}
		\toprule 
		\textbf{Phase} & & Bulk & Powder & 10 wt.\% & 30 wt.\%  & 50 wt.\% & 70 wt.\%\\ \hline 
		\textbf{Fraction} & & 93.4(2)\% & 76.2(2)\% & 88.3(5)\% & 88.9(1)\%  & 88.3(5)\% & 90.0(3)\% \\ \hline
		\begin{tabular}[c]{@{}c@{}}O(I)\\ Pnma\end{tabular} & \begin{tabular}[c]{@{}c@{}}\textbf{a} \\ \textbf{b} \\ \textbf{c} \\ \textbf{V} \end{tabular} & \begin{tabular}[c]{@{}c@{}} 7.505(3) \AA\\  14.78(1) \AA\\  7.759(2) \AA\\  864.8(1) \AA$^3$\end{tabular} & \begin{tabular}[c]{@{}c@{}} 7.545(3) \AA\\  14.77(4) \AA\\ 7.797(2) \AA\\ 869.0(5) \AA$^3$\end{tabular} & \begin{tabular}[c]{@{}c@{}} 7.488(1) \AA\\  14.73(1) \AA\\  7.764(5) \AA\\  856.2(1) \AA$^3$\end{tabular} & \begin{tabular}[c]{@{}c@{}} 7.469(1) \AA\\  14.69(1) \AA\\  7.733(0) \AA\\  849.8(9) \AA$^3$\end{tabular} & \begin{tabular}[c]{@{}c@{}} 7.458(5) \AA\\  14.69(1) \AA\\  7.733(4) \AA\\  847.4(9) \AA$^3$\end{tabular} & \begin{tabular}[c]{@{}c@{}} 7.435(2) \AA\\  14.69(3) \AA\\ 7.769(2) \AA\\  846.5(3) \AA$^3$\end{tabular} \\ \hline
		\textbf{Fraction} & & 6.51(6)\% & 23.4(2)\% & 9.84(7)\% & 10.2(4)\% & 10.0(7)\% & 9.32(5)\% \\ \hline
		\begin{tabular}[c]{@{}c@{}}M\\ P112$_1$/a\end{tabular}  & \begin{tabular}[c]{@{}c@{}}\textbf{a}  \\ \textbf{b} \\ \textbf{c} \\ \textbf{$\gamma$} \\ V \end{tabular} & \begin{tabular}[c]{@{}c@{}} 7.512(8) \AA\\  14.64(1) \AA\\  7.837(7) \AA\\  93.06(4)$^o$\\  860.5(1) \AA$^3$\end{tabular} & \begin{tabular}[c]{@{}c@{}} 7.489(1) \AA\\ 14.73(1) \AA\\ 7.786(5) \AA\\  93.26(7)$^o$\\  857.3(4) \AA$^3$\end{tabular} & \begin{tabular}[c]{@{}c@{}} 7.508(1) \AA\\  14.69(2) \AA\\  7.743(1) \AA\\ 93.09(1)$^o$\\  852.8(1) \AA$^3$\end{tabular} & \begin{tabular}[c]{@{}c@{}} 7.486(2) \AA\\ 14.67(2) \AA\\ 7.720(2) \AA\\ 93.27(1)$^o$\\ 846.4(2) \AA$^3$\end{tabular} & \begin{tabular}[c]{@{}c@{}} 7.445(2) \AA\\  14.65(2) \AA\\  7.764(2) \AA\\  93.03(7)$^o$\\  845.5(3) \AA$^3$\end{tabular} & \begin{tabular}[c]{@{}c@{}} 7.455(1) \AA\\  14.62(0) \AA\\ 7.760(1) \AA\\ 92.9(2)$^o$\\  844.547(3) \AA$^3$\end{tabular} \\  \hline
		\textbf{Fraction} & & 0.12(3)\% & 0.44(2)\% & 1.90(7)\% & 0.84(4)\% & 1.71(8)\% & 0.73(4)\% \\ \hline
		\begin{tabular}[c]{@{}c@{}}5:3\\ P6$_3$/mcm\end{tabular} & \begin{tabular}[c]{@{}c@{}}\textbf{a = b} \\ \textbf{c} \\ \textbf{V} \end{tabular} & \begin{tabular}[c]{@{}c@{}} 8.735(1) \AA\\ 6.287(7) \AA\\ 415.4(2) \AA$^3$\end{tabular} & \begin{tabular}[c]{@{}c@{}} 8.754(2) \AA\\  6.299(7) \AA\\  418.1(3) \AA$^3$\end{tabular} & \begin{tabular}[c]{@{}c@{}} 8.733(4) \AA\\ 6.272(4) \AA\\ 414.3(4) \AA$^3$\end{tabular} & \begin{tabular}[c]{@{}c@{}} 8.726(4) \AA\\  6.279(4) \AA\\ 414.1(3) \AA$^3$\end{tabular}   & \begin{tabular}[c]{@{}c@{}} 8.672(2) \AA\\ 6.255(2) \AA\\ 407.4(2) \AA$^3$\end{tabular}  & \begin{tabular}[c]{@{}c@{}} 8.665(2) \AA\\  6.326(6) \AA\\ 411.4(3) \AA$^3$\end{tabular} \\\hline
		\begin{tabular}[c]{@{}c@{}}U\\ V\\ W\end{tabular} & & \begin{tabular}[c]{@{}c@{}}0.00022\\ -0.00010\\ 0.00135\end{tabular} & \begin{tabular}[c]{@{}c@{}}0.00403\\ -0.00212\\ 0.00351\end{tabular} & \begin{tabular}[c]{@{}c@{}}0.00403\\ -0.00212\\ 0.00351\end{tabular}     & \begin{tabular}[c]{@{}c@{}}0.03307\\ -0.02093\\ 0.00601\end{tabular}     & \begin{tabular}[c]{@{}c@{}}0.02927\\ -0.01955\\ 0.00551\end{tabular} & \begin{tabular}[c]{@{}c@{}}0.00225\\ -0.00164\\ 0.00315\end{tabular}  \\ \hline
		\begin{tabular}[c]{@{}c@{}}\textbf{N$^o$ of}\\\textbf{ Parameters}\end{tabular} & & 51 & 49 & 51 & 50 & 43 & 42 \\ \hline
		\begin{tabular}[c]{@{}c@{}}Rp\\ Rwp\\ Rexp\end{tabular} & & \begin{tabular}[c]{@{}c@{}}5.21\\ 7.11\\ 2.84\end{tabular} & \begin{tabular}[c]{@{}c@{}}3.99\\ 5.32\\ 1.55\end{tabular} & \begin{tabular}[c]{@{}c@{}}6.14\\ 7.80\\ 6.04\end{tabular} & \begin{tabular}[c]{@{}c@{}}5.11\\ 7.31\\ 2.38\end{tabular} & \begin{tabular}[c]{@{}c@{}}6.05\\ 8.29\\ 2.52\end{tabular} & \begin{tabular}[c]{@{}c@{}}2.98\\ 4.18\\ 2.36\end{tabular}\\ \hline
		$\chi ^2$ & & 6.27 & 
		3.52 & 1.67 & 9.45 & 10.8 & 3.16  \\
		\hline \hline                                                     
	\end{tabular}
\end{table*}

\begin{figure}
	\centering
	\includegraphics[width=8.5cm,keepaspectratio]{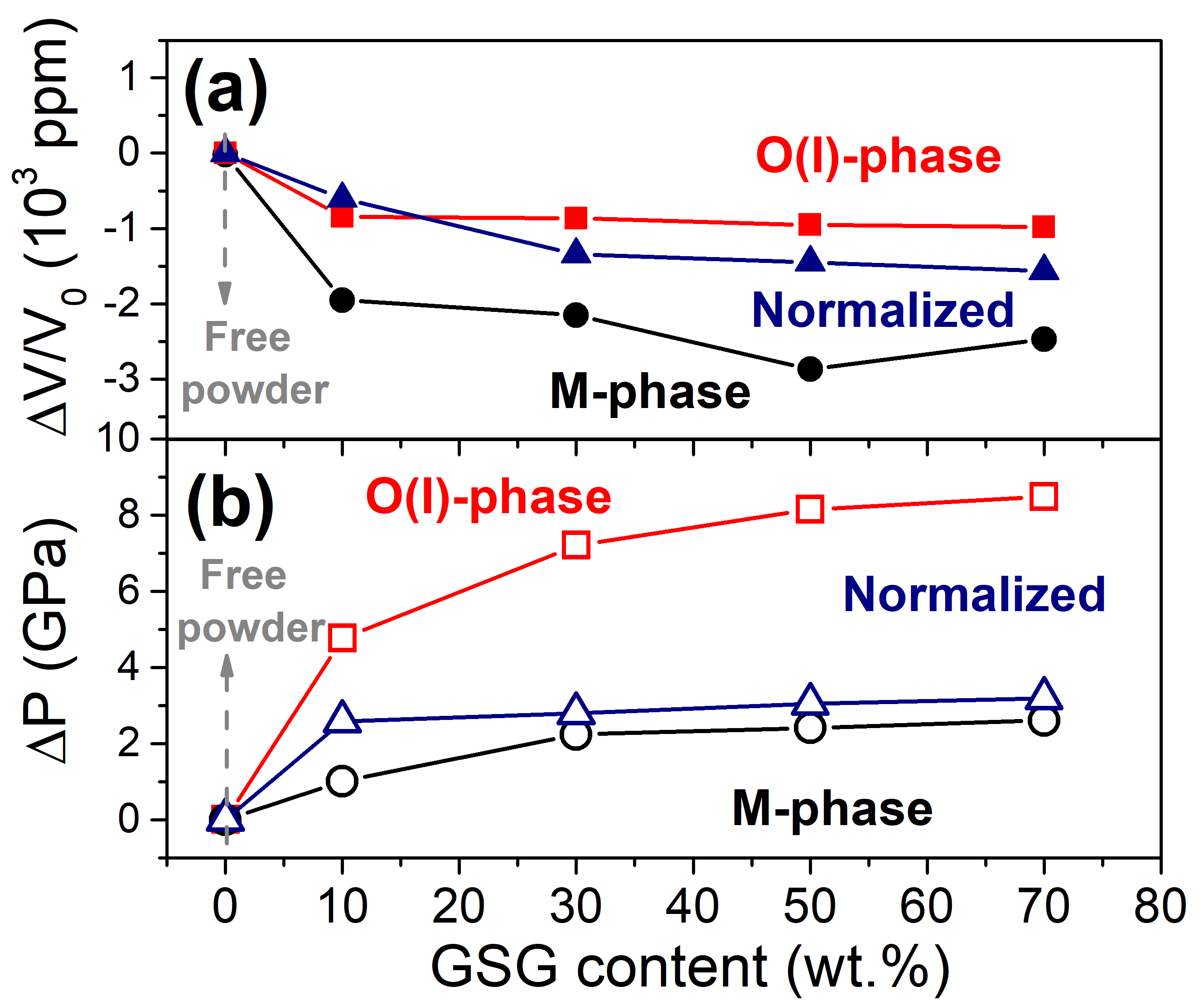}
	\caption{\textbf{(a)} The relative unit cell volume as a function of GSG filler content, where $V_0$ corresponds to the free powder results. The respective hydrostatic pressure applied by PMMA walls in \textbf{(b)} was calculated considering the compressibility values of: $\kappa _{O(I)}$ = 3 TPa$^{-1}$ and  $\kappa _{M}$ = 6 TPa$^{-1}$ from Ref.	\citenum{mudryk2005polymorphism}.}\label{dP_dV}
\end{figure}

Regarding the composite samples patterns, also shown in Fig. \ref{XRD}(c), a shift on the peak positions towards higher 2$\theta$ angles in comparison with the micrometer powder pattern indicates a reduction on the unit cell volume. Besides that, there is a reduction on the intensity of diffracted peaks from secondary M-phase, suggesting that the presence of PMMA is affecting the amount of secondary phases. In order to confirm these, \textit{Rietveld} calculations were performed by considering the resultant lattice parameters and atomic positions from the free powder diffractogram refinement as initial values. Since PMMA is amorphous, there is a large background contribution to the pattern leading to a decrease on \textit{Rietveld} refinement quality; however, the results summarized on Table \ref{PMMA_parameters_XRD} are in fine agreement with previous reports\cite{pecharsky1997phase,misra2006distribution}. The introduction of magnetocaloric powder on the polymeric matrix lead to a significant enhancement on the phase fraction of O(I)-phase from $\sim$76.2\% for the free powder to $\sim$88.9\% in average for the composite samples. Concomitantly, the detection of M-phase reduces from $\sim$22.0\% (free powder) to $\sim$10.0\% (composite). This observation for the reinforced system can be due to an applied pressure on the surface by PMMA. This effect can be attributed to PMMA solidification around the GSG grains that lead to a contraction at the powder boundaries. This is reflected on the unit cell volume reduction $\Delta V/V_0$ of these phases, where $V_0$ correspond to the volume of the free powder, with the increase of GSG content, shown in Fig. \ref{dP_dV}(a). The normalized unit cell volume was obtained considering the O(I), M and 5:3 phase fractions from Table \ref{PMMA_parameters_XRD}, that follows the same behavior as the individual phases. This is mainly due to the extreme sensitivity of M-phase to external parameters, where the PMMA surface in contact with the grains is working as a isotropic pressure cell\cite{pecharsky2001gd5,carvalho2005magnetic}. Mudryk \textit{et. al.} have shown that an applied pressure of 2 GPa is required to induce a complete transition from the M to O(I) structure of Gd$_5$Si$_2$Ge$_2$ polycrystal\cite{mudryk2005polymorphism}. The authors have obtained the isothermal compressibility ($\kappa_T$) of 3 TPa$^{-1}$ and 6 TPa$^{-1}$ for M and O(I) phases, respectively. Considering these values and using the experimental relative reduction on the unit cell volume, it is possible to estimate the isobaric pressure on each structural phase of the grains through the thermodynamic relation: $\kappa_T = - (1/V)(dV/dP)_T$. The calculated values, presented in Fig. \ref{dP_dV}(b), show that the thermoplastic walls force along the M-phase at grain boundaries are estimated to be close to the critical value of 2 GPa for a polycrystalline sample. This might be the reason for an incomplete conversion of M into O(I) phase by the matrix. As a matter of fact, since the present composition have higher Si/Ge ratio than Gd$_5$Si$_2$Ge$_2$ stoichiometry, these values can be overestimated and \textit{in situ} measurements should be performed for accuracy\cite{mudryk2005polymorphism}.

\begin{figure}[htb!]
	\centering
	\includegraphics[width=8.5cm,keepaspectratio]{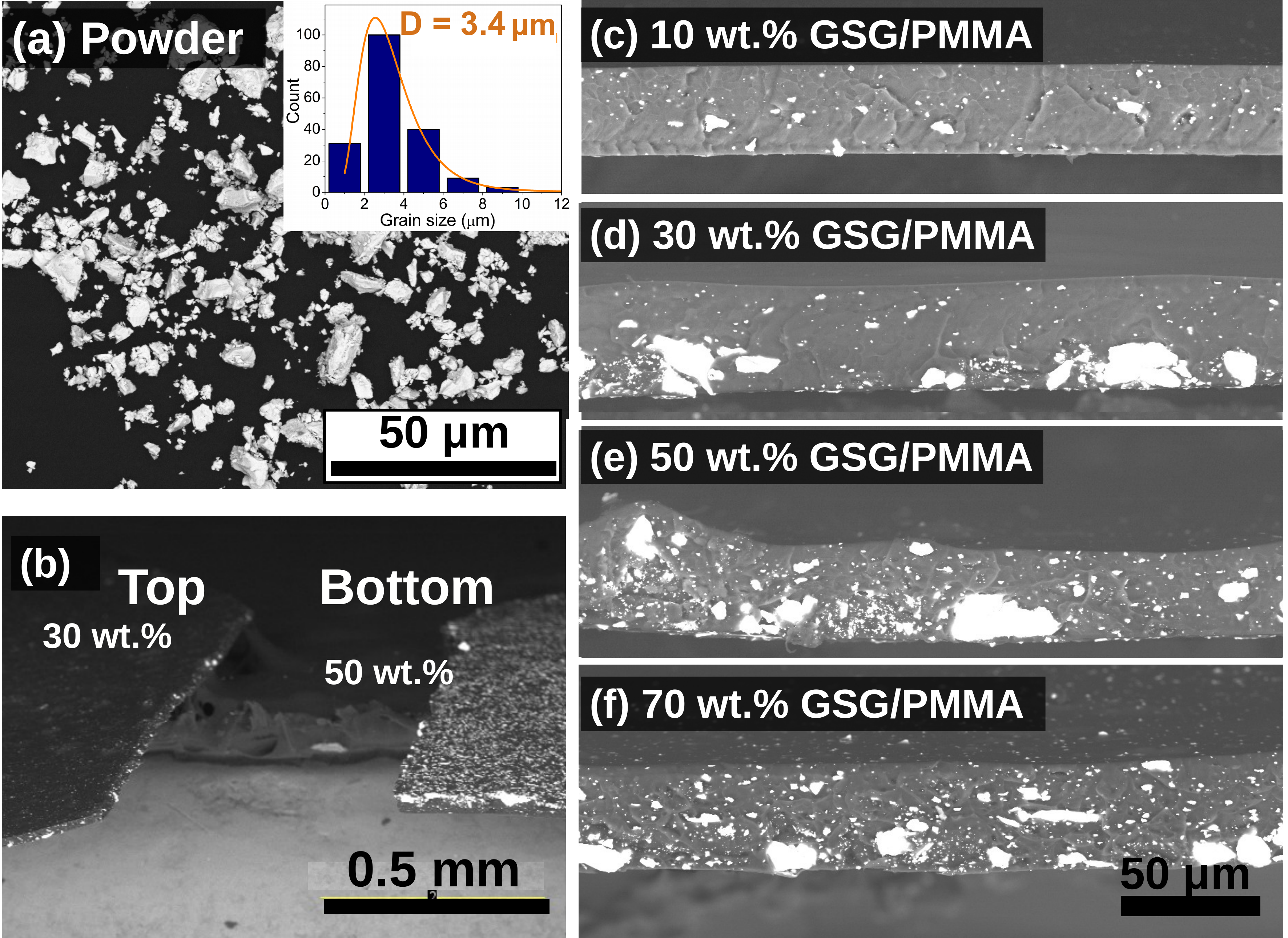}
	\caption{SEM micrographs obtained for \textbf{(a)} sieved powder and its grain size distribution on the \textit{inset}, \textbf{(b)} an image from the particle agglomeration at the bottom of PMMA films surface and cross-section images for all composite samples with \textbf{(c)} 10 wt.\%, \textbf{(d)} 30 wt.\%, \textbf{(e)} 50 wt.\% and \textbf{(f)} 70 wt.\% of filler content, respectively.}\label{SEM}
\end{figure}

SEM images were carried out in order to investigate the efficiency on separating microparticles through sieving the GSG bulk and to verify the particles distribution along the polymeric matrix volume. From the micrograph in Fig. \ref{SEM}(a) it was found that GSG thinner powder displays a log-normal particle size distribution averaging 3.4 $\mu$m, as shown on the \textit{inset}. The presence of bigger grains is due to the fact that they do not present a uniform geometry. Due to gravity, the particles will agglomerate at the bottom, see Fig. \ref{SEM}(b), which led into stress on the PMMA surface and an increase on the curvature. The particle segregation at the bottom is the responsible for a non-regular surface of the films, as can be seen for 50 wt.\% in Fig. \ref{SEM}(b). This evidence is another advantage for application in cooling systems since the thermal contact of the device can be selected at one side and being isolated by the polymer layer at the other\cite{kitanovski2010innovative}. As can be noted also for 50wt.\% composite in Fig. \ref{SEM}(b), the particle distribution along the film surface is quasi-homogeneous. It is known that particles with mean diameters ranging from 2-5 $\mu$m can not be perfectly dispersed onto polymeric matrices even with sonification, stirring and other conventional techniques\cite{guan2008magnetostrictive}.

Furthermore, Fig. \ref{SEM}(c)-(f) shows the cross-section of freeze-fractured composites, where it is possible to notice that the polymer structure grows around the particles, revealing the bonding between particle and PMMA interface. The amplifications on these images does not allow to infer the porosity level of the samples; however, it is possible to notice a few gaps around the grains that suggest higher porosity. These observations justify the intergrain pressure interpretation shown above. As already mentioned, the extreme sensitivity of powder M-phase on the grains boundaries lead to a reduction on the amount of this phase and also its unit cell volume\cite{pecharsky2001gd5}. For comparison purposes, a pure PMMA sample was prepared by following the same procedure and a film with $\sim$ 15 $\mu$m of thickness was obtained. With the addition of 10 wt.\% of GSG microparticles remarkably increases the thickness of the composite to $\sim$ 36 $\mu$m and reach the maximum of $\sim$ 42 $\mu$m for 70wt\% of GSG. This initial characterization will be of great matter for the magnetic and magnetocaloric properties evaluation. 


\subsection{Magnetic characterization}

\begin{figure}[htb!]
	\centering
	\includegraphics[width=8.5cm,keepaspectratio]{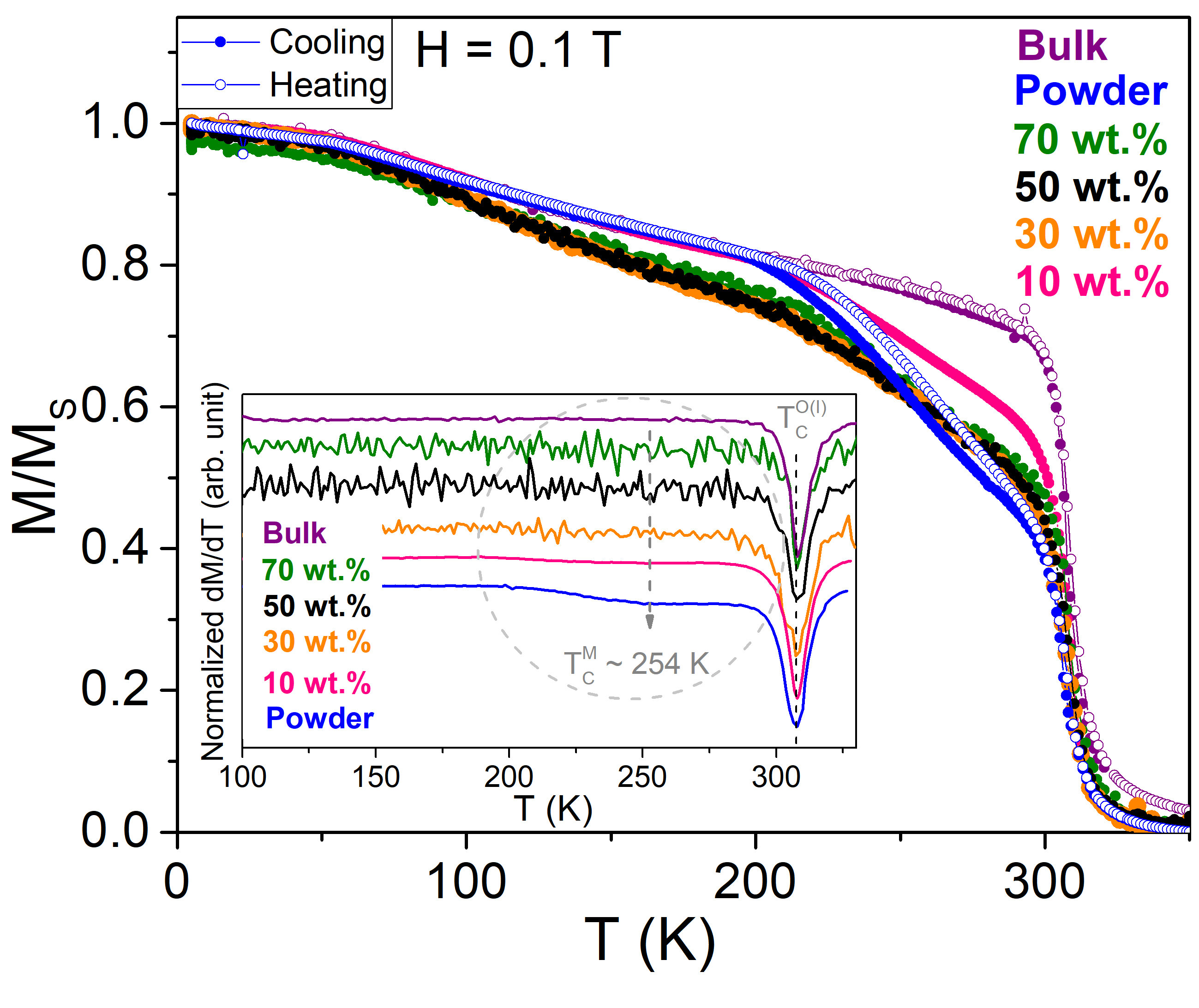}
	\caption{Magnetization as a function of temperature for all samples under an 0.1 T of magnetic field with an amplification for the 10 wt.\% composite showing thermal hysteresis from the M-phase on the \textit{inset}.} \label{MT_MH}
\end{figure}

The normalized M-T curves for the Gd$_5$Si$_{2.4}$Ge$_{1.6}$ in the form of bulk, powder and flexible composite obtained at cooling and heating between 5 and 350 K with an applied magnetic field of 0.1 T are shown in Fig. \ref{MT_MH}(a). The bulk presents a second order ferromagnetic to paramagnetic transition around 308 K are in agreement with previous reports\cite{pecharsky1997phase,misra2006distribution}, as obtained through the $dM/dT$ curves presented on the \textit{inset} of Fig. \ref{MT_MH}(a). Notice that Gd$_5$(Si,Ge)$_4$ compositions that crystallize in a M structure will present a first order magnetic transition (FOMT), simultaneously changing to an O(I) structure at lower temperatures\cite{pecharsky1997phase}. As can be noted in the magnetization profile curves, with the selection of particles with 3.4 $\mu$m mean size, there is a thermal hysteresis ranging from 200 to 300 K that is a consequence on the increase of M-phase content. Such observation is confirmed through the appearance of a bump highlighted at the $dM/dT$ curves\cite{belo2012phase}. It is worth to point out that the reduction of particle size seems to be affecting only the secondary M-phase since there is no shift on T$_C$ for the main O(I) phase, as usually observed in magnetic materials at the micro/nanoscale\cite{pires2015influence,andrade2016magnetocaloric,checca2017phase}. Such effect can be rising from the higher sensitivity of the M structure over O(I) to applied magnetic field, hydrostatic pressure and particle size\cite{carvalho2005magnetic,miller2006complex}. When the micropowder is immersed into the PMMA matrix, there is a slight lift on the magnetization curves at the thermal hysteresis temperature range that lead to a reduction on the bump at the derivative curves on the \textit{inset} of Fig. \ref{MT_MH}(a). As for the O(I) phase temperature transition, there is no shift on T$_C$ due to the presence of non-magnetic PMMA matrix.   



\begin{figure}[htb!]
	\centering
	\includegraphics[width=8.5cm,keepaspectratio]{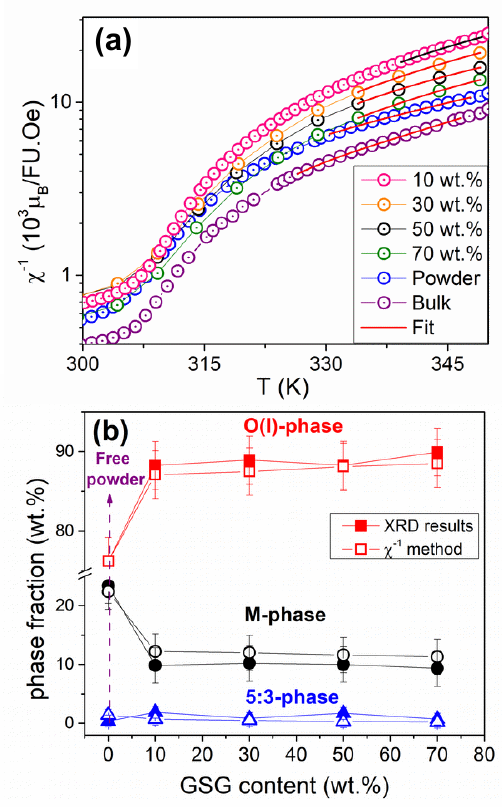}
	\caption{\textbf{(a)} The reciprocal magnetic susceptibility curves with their respective Curie-Weiss law fitting using Eq. (1) and \textbf{(b)} a comparison between the phase fractions obtained from XRD analysis and $\chi ^{-1}$ methods.} \label{susc}
\end{figure}

The strong spin-lattice coupling on the Gd$_5$(Si,Ge)$_4$ family compounds allows the use of its magnetic results to infer the crystallographic phases fractions\cite{belo2012phase}. Above the ferro-paramagnetic temperature transition, the magnetic susceptibility is described by the Curie-Weiss law where we should consider the amount of each phase $x$ on the sample as follows\cite{andrade2018lanthanum}:

\begin{equation}
\chi = \frac{x_{O(I)}C^{O(I)}}{T - \Theta_P^{O(I)}} + \frac{x_{M}C^{M}}{T - \Theta_P^{M}} + \frac{x_{5:3}C^{5:3}}{T - \Theta_P^{5:3}},
\end{equation}

\noindent where C=$\mu _{eff}^2$/3k$_b$ represents the Curie constant and $\Theta_P$, the Curie paramagnetic temperature. In particular, for 10 wt.\% composite, a correction on the magnetization curves due to the diamagnetic contribution from the matrix was performed, also performed in similar composites\cite{tawansi2002effect}. The best curves are shown in Fig. \ref{susc}(a) for all samples with the returned values summarized on Table \ref{Mag_PMMA}. For a better visualization, the amount of each phase is given in Fig. \ref{susc}(b) obtained from $\chi^{-1}$ method and XRD results for comparison purposes. As can be noted, there is no significant change on the values obtained through the different  methods, since they are within the error. Most notably, the magnetic analysis seems to detect a slightly higher content of M-phase than XRD calculations which might be related to the higher sensitivity of $\chi^{-1}$ method. These findings confirm the assumption that PMMA polymeric matrix works as a pressure cell on the grains surface and, thus, weakens the effects of secondary M-phase. Indeed, further evaluation on the MCE results will also reveal the strong influence of the non-magnetic thermoplastic on the magnetocaloric powder. Furthermore, the $\mu_{eff}$ value for this phase is below the expected 7.94 $\mu_B$ which has direct consequences on the MCE features of GSG magnetic material\cite{carvalho2005magnetic}. Furthermore, $\theta _P$ and $\mu_{eff}$ are in fine agreement with the obtained for pure powder and the theoretically expected\cite{belo2012phase,roger2006structural}. Although there is a reduction on $\mu _{eff}$ values, they are still close to theoretically expected 7.94 $\mu_B$ for Gd$^{3+}$ ions\cite{belo2012phase}. Therewith, the presence of PMMA does not affect the intrinsic magnetic features of the 3.4 $\mu$m powder. In another words, the polymeric matrix is acting as an external agent on the grains in form of hydrostatic pressure that have been shown to have great influence on the M-phase\cite{miller2006complex,pecharsky2003giant,mudryk2005polymorphism}. 
 
\begin{table}[]
	\centering
	\caption{Results obtained from the Curie-Weiss law: O(I), M and 5:3 phases calculated values of paramagnetic temperature ($\Theta_P$), the effective moment ($\mu _{eff}$). Important magnetic parameters are also placed: the magnetization at 5 K and 5 T for the composite samples (i.e, taking into account the weight of filler and polymer) and the saturation magnetization ($\mu_{sat}$) obtained considering the grains weight fraction.}\label{Mag_PMMA}
	\begin{tabular}{cccccc}
		\toprule
		\textbf{Sample}  & \textbf{Phase} &  \begin{tabular}[c]{@{}c@{}}$\Theta _P$\\ (K)\end{tabular} & \begin{tabular}[c]{@{}c@{}}$\mu _{eff}$\\ \tiny({$\mu _B /Gd^{3+}$})\end{tabular} & \begin{tabular}[c]{@{}c@{}}M(5K,5T)\\ ($\mu _B$/FU)\end{tabular}  & \begin{tabular}[c]{@{}c@{}}$\mu_{sat}$\\ ({$\mu _B /Gd^{3+}$})\end{tabular} \\ \hline \hline
		\textbf{Bulk}  & \begin{tabular}[c]{@{}c@{}}O(I) \\ M \\ 5:3 \end{tabular} & \begin{tabular}[c]{@{}c@{}}304(7)\\ 302(9)\\164(3)\end{tabular}  &   \begin{tabular}[c]{@{}c@{}}7.93(5)\\ 7.66(4)\\8.13(5)\end{tabular}  & 34.7(5)  & \begin{tabular}[c]{@{}c@{}}7.83(6)\\\end{tabular}   \\ \hline
		\begin{tabular}[c]{@{}c@{}}\textbf{Powder}\end{tabular} & \begin{tabular}[c]{@{}c@{}}O(I) \\M \\ 5:3 \end{tabular}  &  \begin{tabular}[c]{@{}c@{}}310(4)\\ 293(3)\\186(4)\end{tabular}  & \begin{tabular}[c]{@{}c@{}}7.92(5)\\ 7.58(7)\\8.16(6)\end{tabular}  & 30.6(2) & \begin{tabular}[c]{@{}c@{}} 7.54(8)\\\end{tabular}\\ 	\hline
		\textbf{70 wt.\%}  &  \begin{tabular}[c]{@{}c@{}} O(I)\\ M \\  5:3 \end{tabular} &   \begin{tabular}[c]{@{}c@{}}310(3)\\ 291(4) \\ 186(5) \end{tabular} &  \begin{tabular}[c]{@{}c@{}}7.92(2)\\ 7.45(7) \\ 8.03(2) \end{tabular}  & 22.03(4) & 6.96(2) \\ \hline
		\textbf{50 wt.\%}  &  \begin{tabular}[c]{@{}c@{}} O(I)\\  M \\  5:3 \end{tabular} &   \begin{tabular}[c]{@{}c@{}}310(5)\\ 296(6) \\ 186(5) \end{tabular} & \begin{tabular}[c]{@{}c@{}}7.94(2)\\ 7.51(5) \\ 8.20(3)\end{tabular} & 15.5(5) & 6.99(6) \\ \hline
		\textbf{30 wt.\%}  & \begin{tabular}[c]{@{}c@{}} O(I)\\  M \\  5:3 \end{tabular} &   \begin{tabular}[c]{@{}c@{}}308(3)\\ 289(5) \\ 183(3)\end{tabular} &  \begin{tabular}[c]{@{}c@{}}7.93(4)\\ 7.59(3) \\ 8.03(6) \end{tabular} & 9.03(1) & 6.67(2)  \\ \hline 
		\textbf{10 wt.\%}  & \begin{tabular}[c]{@{}c@{}} O(I)\\ M \\  5:3\end{tabular} &   \begin{tabular}[c]{@{}c@{}}308(5)\\ 294(5) \\ 184(6)\end{tabular} & \begin{tabular}[c]{@{}c@{}}7.89(6)\\ 7.91(4) \\ 8.09(3)\end{tabular} & 3.04(1) & 6.14(1) \\  \hline \hline
	\end{tabular}
\end{table}

Nevertheless, magnetization data at 5 K with an applied magnetic field up to 5 T for the bulk, powder and composite samples were acquired for rating their M(5K,5T) value. Using this data it was also possible to extract their saturation magnetization ($\mu_{sat}$) values through an extrapolation on the M \textit{versus} 1/H, with the values are summarized in Table \ref{Mag_PMMA}. As can be noted, there is a reduction on M(5K,5T) and $\mu_{sat}$ values from the bulk to the powder sample that is related to dimensionality reduction and from the 5:3-phase AFM ordering - the T$_N$ of this phase is report to be 75 K\cite{roger2006structural,belo2012phase,hunagund2018investigating}. As for the composite samples, taking into account the contribution from all the system components, \textit{i.e.}, m$_{GSG} + $m$_{PMMA}$, we obtain the M(5K,5T) quantity that reduces as the amount of magnetic material decreases. This is a direct consequence on the dilution of magnetic powder along the non-magnetic PMMA matrix, as observed in previous magnetocaloric composites\cite{zhang2015mechanical}. The $\mu_{sat}$ values, however, were calculated by considering only the weight fraction of active magnetic material and the contribution from each crystallographic phase obtained through $\chi^{-1}$ method. In a similar way, there is a loss on the $\mu_{sat}$ values for lower GSG content composites that can be due to dilution effects or from possible evaporation losses during synthesis\cite{ramprasad2004magnetic}. For this reason, the obtained values are within the error which corroborates with previous assumptions that the presence of a non magnetic polymer does not affect the intrinsic properties of magnetocaloric filler. Furthermore, the saturation is reached at lower intensities of magnetic field for the composite samples with lower filler contents and it starts to behave as the free powder with increasing the amount of magnetic material. Such behavior can raise from interparticle short range interactions with associated low level of deformation, as observed in soft magnetic elastomers, that can also contribute to this fast magnetization on the composite samples\cite{stolbov2011modelling,stepanov2007effect}.

\section{Magnetocaloric Effect}

\begin{figure}
	\centering
	\includegraphics[width=8.5cm,keepaspectratio]{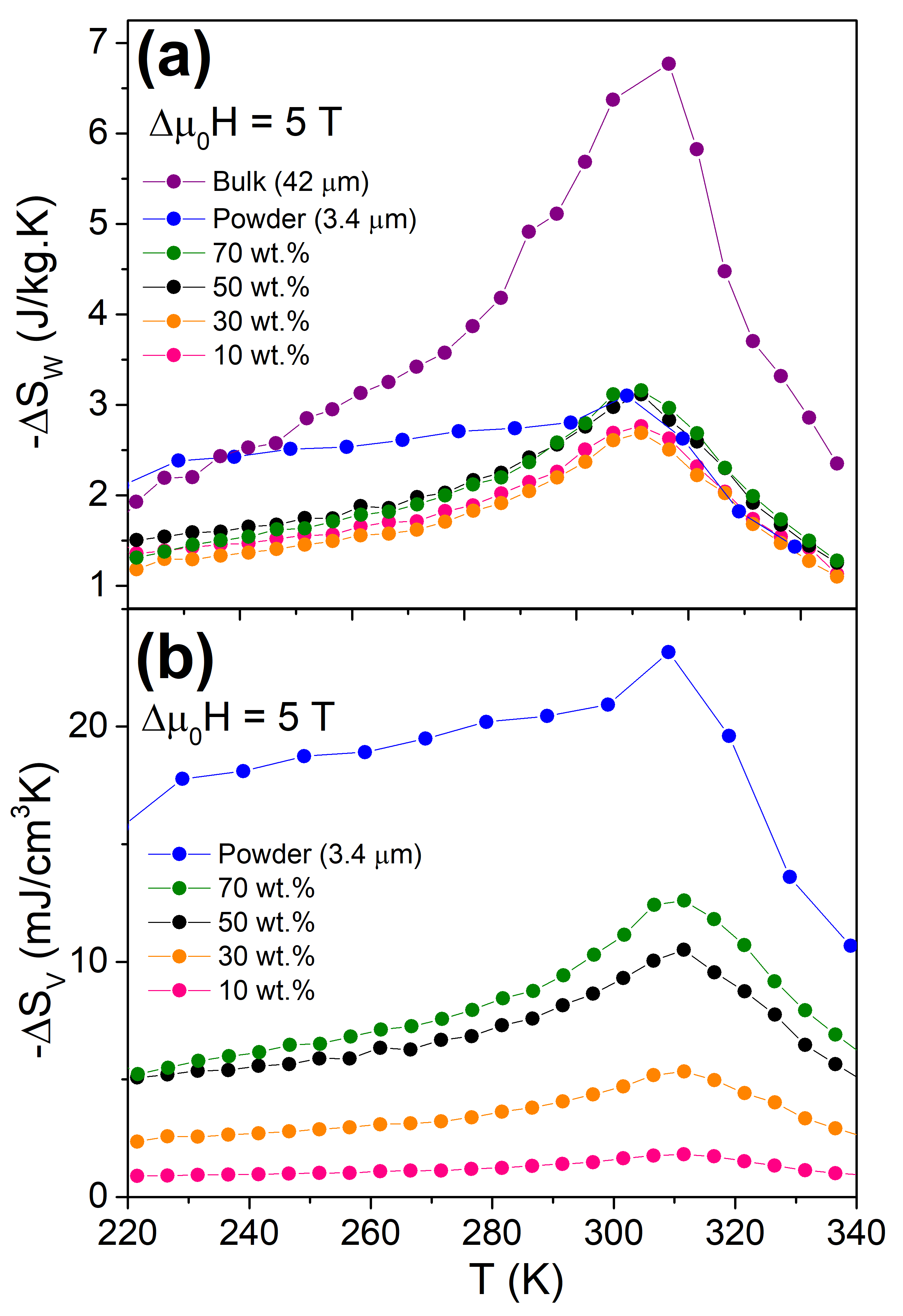}
	\caption{Temperature dependence of the entropy change ($\Delta S$) obtained using Eq. 2 for an applied magnetic field of 5 T considering the \textbf{(a)} weight of magnetic active material and \textbf{(b)} volumetric magnetization values.} \label{MCE_PMMA}
\end{figure}

The MCE is a well known phenomenon where the thermodynamic properties of a ferromagnetic material can be reversible varied through the application of a magnetic field. The thermal variations can be observed in two basic processes: \textit{i)} adiabatic, that leads into changes in temperature ($\Delta T$), and \textit{ii)} isothermal, that is quantified by an entropy change ($\Delta S$). For the present work, the MCE evaluation were performed through magnetization isothermal measurements obtained from 215 K to 350 K by increasing and decreasing the applied magnetic field up to 5 T, given on SID Fig. S3 and S4. It is worth to point out that, due to the magnetic irreversibility, the samples were warmed up to 330 K between each isothermal measurement. Therefore, through the magnetization map M(T,H), the magnetocaloric potential $\Delta S$ was calculated by using the integrated Maxwell relation $\Delta S (H,\Delta T) = \int\limits_{0}^{H} \left(\frac{\partial M(T,H)}{\partial T} \right) dH$\cite{de2010theoretical}. For the calculations, first it was considered only the contribution from the filler, \textit{i.e.}, the weight of magnetic active material, denoted as $\Delta S_W$. As can be noted, there is a reduction on the saturation magnetization as filler content decreases with the diamagnetic contribution from PMMA being more evident for 10 wt.\% sample, as mentioned above. The obtained $\Delta S_W(T)$ curves for powder and composite samples are depicted in Fig. \ref{MCE_PMMA}(a) for $\Delta \mu_0 H = 5 T$. The $\Delta S_W$ curves for the composites follow a $\lambda$-shape, typical of SOMT, and its maximum values raise as the applied magnetic field increases\cite{belo2012phase}. From these curves, the important MCE parameters were extracted as can be seen on Table \ref{MCE_values}.
 


\begin{table}
	\centering
	\caption[Summary on the MCE properties of PMMA/GSG composites.]{Magnetocaloric properties for all the synthesized samples obtained for an applied magnetic field of 5 T calculated considering the sample weight and volume. The RCP values were calculated considering the full width at half maximum (FWHM)\cite{de2010theoretical}.}\label{MCE_values}
	\begin{center}
		\begin{tabular}{ccccc}
			\toprule
			Sample         & \begin{tabular}[c]{@{}c@{}}$\Delta S_W^{max}$\\ (J/kg.K)\end{tabular} & \begin{tabular}[c]{@{}c@{}}$\Delta S_V^{max}$\\ (J/kg.K)\end{tabular} & \begin{tabular}[c]{@{}c@{}}$\delta T_{FWHM}$\\ (K)\end{tabular} & \begin{tabular}[c]{@{}c@{}}$RCP_{FWHM}$\\ (J/kg)\end{tabular} \\ \hline \hline
			Bulk           & 6.04  &               50.4                                    & 23.4                                                     & 69.4                                                       \\
			Powder         & 3.09  &        23.2                                           & 23.8                                                     & 33.1                                                       \\
			70 wt.\% & 2.64  &   12.5                                                & 26.4                                                     & 26.2                                                       \\
			50 wt.\% & 1.14         &    10.4                                        & 25.2                                                     & 10.6                                                       \\
			30 wt.\% & 0.46         &     5.35                                       & 25.2                                                     & 4.50                                                       \\
			10 wt.\% & 0.99     &    1.84                                            & 24.7                                                     & 8.84                                                       \\ \toprule
		\end{tabular}
	\end{center}
\end{table}

Concerning application goals, the volumetric entropy change $\Delta S_V$ is rather informative for device engineering than the mass one\cite{zhang2015mechanical}. The magnetic material density ($\rho$) was considered to be 7.45 g/cm$^3$, in agreement with reported by Gschneidner and Pecharsky\cite{gschneidner2000nonpareil}. As for the composite samples, the density was obtained through a carefully measurement on the film area and considering the thickness obtained by SEM cross-section imaging. It is worth to point out that the acquired density for the systems with 10 and 30 wt.\% of magnetic material are below the expected and might be the reason for loss on $\Delta S^{max}_M$, shown in Fig. \ref{MCE_PMMA}(b). As previously mentioned, this can also be a consequence of a non perfect homogeneous distribution along the film volume, that cannot be assured during casting\cite{jesson2012interface}, and from possible evaporation losses during synthesis\cite{ramprasad2004magnetic}. A close composition Gd$_5$Si$_{2.5}$Ge$_{1.5}$ have been reported to present a maximum entropy of 70.7 mJ/cm$^3$K at 313 K\cite{gschneidner2000magnetocaloric}. Taking into account the formation of secondary M-phase and eutectic 5:3 phase, the obtained value of $\sim$ 50 mJ/cm$^3$K for 42 $\mu$m are close to the expected\cite{gschneidnerjr2005recent}. For the smaller grains, it drastically decreases to $\sim$ 25 mJ/cm$^3$K with the same profile as shown in Fig. \ref{MCE_PMMA}(a). Although the effect of M-phase is blocked by the polymeric matrix, the presence of non-magnetic material diminish the system magnetocaloric response. However, the values of 5-10 mJ/cm$^3$K are in the range for applications in micro-cooling devices for pump systems\cite{kitanovski2010innovative}.


Regarding the $\Delta S_W$ and $\Delta S_V$ curves features, the bump from 220 to 300 K is not observed for the GSG powder when is blended with PMMA thermoplastic. As already illustrated, the particles are confined on the polymeric matrix that can be acting as a pressure cell on the grain boundaries. It is known that Gd$_5$(Si,Ge)$_4$ is highly sensitive to internal and external stimuli\cite{pecharsky2001gd5}. The thermal expansion of the filler, monoclinic Gd$_5$Si$_2$Ge$_2$ stoichiometry presents an anomalous behavior on it volumetric thermal expansion ($\gamma$) that reaches a maximum $\sim$1.46$\times$10$^{-2}$ ppm/K for a single-crystal\cite{han2002thermal}, being the mechanism responsible for the GMCE on this material. Carvalho \textit{et. al.} have already shown that by applying a hydrostatic pressure of 0.1 GPa reduces the $\Delta S_{max}$ of Gd$_5$Si$_2$Ge$_2$ compound in $\sim$23\% and eventually vanishes after 0.6 GPa due to a suppression of the FOMT\cite{carvalho2005magnetic}. Given this, we can assume that during heating, the thermal expansion of the M-phase changing to an O(I) by the sliding of the pseudo-blocks on the crystal structure is limited by PMMA surface around the grains - since the polymer present a lower thermal expansion\cite{porter2000thermal}. Besides that, it would be interesting to perform studies on temperature cycling of the composite samples to observe if the polymer deformation is reversible, which could allow the structural change on the magnetic material. Since there is particles with a broad particle size distribution and PMMA thermal expansion is anisotropic, there is no simple solution to describe the mechanism behind this effect to estimate the applied pressure from the matrix to the grains edges. For this reason, we can only assume that is above the 0.6 GPa observed on M Gd$_5$Si$_2$Ge$_2$ single crystals. Nonetheless, these findings reveal the interplay between mechanical and magnetocaloric properties that can be used to tune the best material features to produce multifunctional devices.

\section{Conclusions}

The results here presented shows that solvent casting is a suitable technique for the implementation of 10, 30, 50 and 70 weight fraction of 3.4$\mu$m Gd$_5$Si$_{2.4}$Ge$_{1.6}$ particles in non-magnetic PMMA. The micropowder was obtained through sieving and the reduction of particle size have intensified the effect of deformity on the grain boundaries, leading to a detection of $\sim$23\% of deformed M-phase. Although there are no changes on the magnetic nature of the microparticles when blended with the thermoplastic, a slight reduction on the amount of the secondary M-phase is observed. The saturation magnetization at 5 K reveal that the composites magnetic response are ruled by the particle density, being bigger for larger amounts of magnetic filler. This is reflected on the MCE results where the interface GSG/PMMA interaction seems to weaken the contribution from M-phase on the $\Delta S$ curves with the absence of the bump observed for the free powder. Further investigations on thermal and mechanical properties are required for a fully understand on the mechanism of GSG/PMMA composites during thermal cycles. Furthermore, these observations indicates that the system here present can allow several applications such as energy harvesting, microfluidic system and magnetic refrigeration\cite{hamann2017high,harstad2017enhancement,zhang2015mechanical,crossley2015new}. In addition, the findings presented in this work open new avenues on the next generation of the magnetocaloric effect\cite{kitanovski2010innovative}.


\section*{acknowledgement}
	
	This work is funded by FEDER funds through the COMPETE 2020 Programme and National Funds throught FCT - Portuguese Foundation for Science and Technology under the project UID/NAN/50024/2013 and by NECL with the project NORTE-01-0145-FEDER-022096. This work was also supported by the European Union’s Horizon -2020 research and innovation program under the Marie Sklodowska-Curie Grant Agreement No. 734801. VMA thanks the CNPq for the Grant No. 203180/2014-3. J.H. Belo would like to thank CICECO-Aveiro Institute of Materials, POCI-01-0145-FEDER-007679 (FCT Ref. UID /CTM /50011/2013), financed by national funds through the FCT/MEC and when appropriate co-financed by FEDER under the PT2020 Partnership Agreement.
	


%

\end{document}